\documentclass[preprint,double-spaced,aps,apssymb,showkeys,showpacs]{revtex4-1}
\usepackage{amsmath,amssymb}
\usepackage{graphicx}% Include figure files
\usepackage{dcolumn}% Align table columns on decimal point
\usepackage{bm}% bold math
\usepackage{subfigure}
\usepackage{color}

\newcommand{\ber}{\begin{eqnarray}}
\newcommand{\eer}{\end{eqnarray}}
\newcommand{\bea}{\begin{equation}}
\newcommand{\eea}{\end{equation}}

\begin{document}

\title{Equilibrium of a Brownian particle in an inhomogeneous medium: An alternative approach }
%\title{Equilibrium of a Brownian particle under damping induced inhomogeneity: An alternative approach }
\author{Avik Biswas}
\email{avik@iiserb.ac.in}
 \affiliation{Indian Institute of Science Education and Research, Bhopal, India}

\author{A. Bhattacharyay}
\email{a.bhattacharyay@iiserpune.ac.in}
 \affiliation{Indian Institute of Science Education and Research, Pune, India}

\date{\today}

\begin{abstract}
We look at the equilibrium of a Brownian particle in an inhomogeneous space following the alternative approach proposed in ref.\cite{genfdt}. We consider a coordinate dependent damping that makes the  stochastic dynamics the one with multiplicative noise. Here we show that the mapping to an additive noise gives the equilibrium distribution of the generalized Langevin dynamics of a particle with mass. The procedure does not need inclusion of any ad hoc current cancelling term in the Langevin dynamics. The result shows a modified Maxwell-Boltzmann distribution with a damping dependent amplitude.
\end{abstract}
\pacs{05.10.Gg, 05.40.-a, 05.40.Jc, 05.70.-a}% PACS, the Physics and Astronomy
                             % Classification Scheme.
\keywords{Fluctuation-dissipation, canonical distribution, multiplicative noise}%Use showkeys class option if keyword
                              %display desired
\maketitle
\par 
A Brownian particle (BP) in a homogeneous heat bath equilibrates at the minimum of a potential with a Boltzmann-distribution of positions and a Maxwell-distribution of velocity (which from now on we would refer to as canonical distribution in general). The strength of the Gaussian stochastic noise acting on the BP in equilibrium with a heat bath is a function of the damping constant and temperature of the bath. The noise strength is given by $\sqrt{2\Gamma k_B T}$ where $\Gamma$ is the damping constant, $k_B$ is the Boltzmann constant and $T$ is temperature. The stochastic force strength written in this way as a function of damping (response) is a consequence of the fluctuation-dissipation relation to ensure equilibrium with Boltzmann distribution of positions \cite{lem,van,ris}. For a BP in an inhomogeneous heat bath, the Langevin dynamics is characterized by a multiplicative noise (coordinate dependent damping). Conventionally, also in such cases, one uses a straight forward generalization of the stochastic force strength to $\sqrt{2\Gamma(x)k_BT}$ to ensures a canonical distribution in equilibrium. This, however, is in direct conflict with the other more important requirement of having a zero current in equilibrium in an inhomogeneous space \cite{genfdt}. In the standard way of solving this problem using Stratonovich or It\^o conventions, an ad hoc adjustment is customarily done to the Langevin dynamics with multiplicative noise to cancel out the unwanted current. 
\par
{ A lot of recent work focussing on anomalous diffusion and weak ergodicity breaking under mostly non-equilibrium situations is worth mentioning in this context, because, those in general involve a heterogeneous medium. Considering space dependent diffusivity, Metzler and co-workers have shown various situations of  non-ergodic and anomalously diffusive regimes and have put forward alternative approaches to the existing theories \cite{metz1,metz2,metz3}. Considering spatially non-uniform temperature Fuli\'nski has demonstrated a class of anomalous diffusion \cite{ful1} and has also considered G-process with weak ergodicity breaking which is an interesting work on understanding transport in strongly non-equilibrium systems \cite{ful2}. A random diffusion model proposed by Massignan et. al. is also an interesting work in the regime of non-ergodicity \cite{mass}. In the present paper, however, we are strictly interested in the equilibrium in an inhomogeneous medium. We, therefore, cannot consider a spatially or temporally varying temperature of the system. Here, we would be considering a general spatially varying damping and would consider a different damping dependent stochastic force strength than the one mentioned above to ensure equilibrium.}  
\par
There exists proof of having Boltzmann distribution in equilibrium based on maximizing entropy of a system at thermodynamic limit {(see, for example, the section {\it Alternative derivation of the canonical distribution} in chap.6 of the ref.\cite{reif})}. There, however, does not exist any proof for a mesoscopic system to have a canonical distribution in equilibrium with a heat bath in general. { For a BP in a homogeneous medium the Boltzmann distribution is imposed by considering the noise strength as $\sqrt{2\Gamma k_B T}$ and the result is experimentally verified. We argue that, the same distribution which works at thermodynamic limit is also seen to work for a single particle because of the homogeneity of space. It does not mean that the same distribution would work for all the system sizes between the thermodynamic limit and a single particle, in general, if the homogeneity of space is not there.}
\par
{A system at thermodynamic limit as well as a BP in a homogeneous heat bath for which the homogeneity of space is broken solely by the presence of a conservative force field the equilibrium distribution is Boltzmann. Thus, a reflection of the inhomogeneity of space is essentially present in the equilibrium probability distribution. One can argue, therefore, that, if there are other sources causing inhomogeneity of space over which the dynamics takes place, a reflection of that other source should also be there on the ensuing equilibrium distribution. This should be more so for the necessary restriction of zero currents in equilibrium over an inhomogeneous space. Interestingly, in the present context, the homogeneity breaking agent is the dissipative force arising from the interactions with the bath degrees of freedom through geometry (we assume). This is a dissipative interaction in the presence of which equilibrium can in principle be achieved for a mesoscopic system. Since we expect a deviation from the Boltzmann distribution, in what follows, we do not consider the presence of this distribution to be a condition of equilibrium. We stick to the condition of no current in equilibrium as the fundamental condition and look for the emerging distribution.} 
\par
{For a BP to ever equilibrate in an inhomogeneous space, the inhomogeneous space must be of finite extension. Secondly, for such an inhomogeneous space to exist and to be maintained, there in general cannot be a recurring energy cost involved. Note that, the conformation space of a complex molecule with conformation dependent damping can be an example of such an inhomogeneous space. Such conformation spaces of a polymer or a colloid or any other complex molecule in contact with a homogeneous heat bath can be a finite space with inhomogeneities. At large times, after the system has come to equilibrium with the heat bath, it would sample its inhomogeneous conformation space solely being driven by equilibrium distributions. The results shown in the present paper would, in general, be applicable to such complex mesoscopic systems with conformation dependent damping.} 
\par
{
In the present paper, we are giving some results of a generalized (with mass) Langevin dynamics of a Brownian particle in an inhomogeneous space. In this work we use the modified stochastic noise strength $\Gamma(x)\sqrt{2mk_BT/<\Gamma>}$ as proposed by us in ref.\cite{genfdt} instead of the conventionally adopted $\sqrt{2m\Gamma(x)k_BT}$ form. The choice of this stochastic noise form is entirely based on the demand of zero current as has been shown in ref.\cite{genfdt}. With this form of the stochastic noise, we show that the dynamics can be mapped to an additive noise problem and we solve this model to get the probability distribution without needing any convention. We numerically integrate the actual model with multiplicative noise as well as the additive noise form of it to show the same probability distribution as obtained analytically. The model with multiplicative noise has been numerically integrated following an It\^o convention without adding any ad hoc term. We claim this amplitude modulated probability distribution to be the equilibrium probability distribution of the system. The most striking result of the present analysis is the identification of the fact that, the equilibrium probability distribution of a free BP in an inhomogeneous space is not uniform, rather, is a function of the inhomogeneities over space to keep the local velocity distribution as the one of a zero mean.
} 
\par
In what follows, we will present a brief discussion of our previous calculations and results of the over damped system (standard Langevin dynamics) first. Then, we would explain the generalized model with mass term and the mapping to additive noise form to solve it without any convention. Following that, We would present numerical results related to various inhomogeneity profiles. Since, these are numerical results based on Langevin dynamics, its more of a systematic cross-check of our analytical results than an actual numerical verification which would require an atomistic simulation. We end this paper with a discussion on the perceived implication of the present results for the problem of protein folding.
 
\section {Over-damped Model: A review}
Let us have a quick review of the equilibrium dynamics and probability distribution of an over-damped model presented in ref.\cite{genfdt}. The Langevin dynamics of a BP in 1D with multiplicative noise is given by
\bea
\frac{\partial x}{\partial t} = \frac{F(x)}{\Gamma(x)} + \frac{g(x)}{\Gamma(x)}\eta(t),
\eea 
where $F(x)=-\frac{\partial V(x)}{\partial x}$ and $\Gamma(x)$ is a space dependent damping constant and $\eta(t)$ is a Gaussian white noise of unit strength. {In this kind of a stochastic problem the mean value theorem does not work for the integrals involving the stochastic term due to its discontinuous nature. One generally adopts a convention of fixing the stochastic force strength within the small intervals of time $dt$ while doing time integrations. Let us understand this procedure taking help of a generalized Stratonovich convention adopted by Lau and Lubensky}.
\par
The generalized Stratonovich convention as adopted by Lau and Lubensky \cite{lau} deals with the dynamics in the following way. Consider $c$ to be the fraction of the infinitesimal time interval $dt$. They consider the noise strength $\Gamma(x)$ at $c$ within every $dt$ to get the stochastic contribution to the evolution of $x$. As usual, the noise strength is taken to be $g(x)=\sqrt{2\Gamma(x)k_BT}$. This requires one to add an extra term to the right hand side of Eq.1 to make an unwanted probability current in equilibrium vanish. The term added is $(1-c)g_1(x)\partial g_1(x)/\partial x$ where $g_1(x)=g(x)/\Gamma(x)$. For $c=0$ its an It\^o convention, when $c=1/2$, the convention is a standard Stratonovich one and otherwise its the generalized Stratonovich convention of Lau and Lubensky \cite{lau}.
\par
Note that, irrespective of conventions adopted, one must add at least the term $g_1(x)\partial g_1(x)/\partial x$ to the r.h.s. of the above equation to cancel out an equilibrium probability current in an ad hoc manner. This is a consequence of straightforwardly generalizing the noise strength taken over a homogeneous space to an inhomogeneous space like $g(x)=\sqrt{2\Gamma(x)k_BT}$ to ensure a canonical position distribution. This is exactly where the effectively considered postulates of the theory - in equilibrium the probability distribution must be canonical and there cannot be any probability current in equilibrium in an inhomogeneous space are at conflict.
\par
In our proposal ref.\cite{genfdt}, we have suggested that, one cannot base a theory on conflicting postulates and must drop the less fundamental one which, at present, is the demand of the canonical probability distribution. Sticking to the demand of no probability current in equilibrium in an inhomogeneous space, we have already derived in ref.\cite{genfdt} that the noise strength is $g(x)=\Gamma(x)\sqrt{2k_BT/\left <\Gamma(x)\right >}$. This is not a straightforward generalization of the form $g=\sqrt{2\Gamma k_BT}$. As a proof of the fact that our proposed noise strength ensures equilibrium one can look at the exact mapping of the dynamics in the inhomogeneous space to the equilibrium dynamics in a homogeneous space. 
\par
Equilibrium probability distribution of a BP over a homogeneous space (constant damping) in the presence of a global force field having the dynamics of the form
\bea
\frac{\partial x}{\partial t} = \frac{F(x)}{\Gamma} + \frac{g}{\Gamma}\eta(t),
\eea
is given by
\bea
P(x)=N\exp\left ({\frac{2}{g_1^2}\int{dx \frac{F(x)}{\Gamma}}}\right ),
\eea
where $g_1=g/\Gamma$. This expression results in the Boltzmann distribution $P(x)=N\exp({-V(x)/k_BT})$ for $g=\sqrt{2\Gamma k_BT}$, which can be checked easily where $N$ is the normalizing constant. 
\par
Now, the Langevin dynamics over an inhomogeneous space with the noise strength as derived in ref.\cite{genfdt} is of the form
\bea
\frac{\partial x}{\partial t} = \frac{F(x)}{\Gamma(x)} + \sqrt{\frac{2k_BT}{\left <\Gamma(x)\right >}}\eta(t),
\eea 
which is exactly of the form shown in Eq.2 enabling us to readily write down the equilibrium probability distribution 
\bea
P(x)=N\exp\left ({\frac{\left <\Gamma(x)\right >}{k_BT}\int{dx \frac{F(x)}{\Gamma(x)}}}\right ).
\eea
{This is the form we have arrived at in ref.\cite{genfdt} on the basis of demanding no current in an over-damped system.} 
On the other hand, it can be easily checked using this equilibrium probability distribution that average current in equilibrium
\bea
\left <\frac{\partial x}{\partial t}\right > = \left <\frac{F(x)}{\Gamma(x)}\right > = \int_0^0{dP(x)} =0.
\eea
\par
There is no need to add any ad hoc term to the Langevin dynamics to make the current vanish. Note that, the presence of the weighted average of damping $\left <\Gamma(x)\right >$ in the expressions would demand that the system has already seen the whole inhomogeneous space before equilibrating. Such an equilibrium would require the system to see the whole inhomogeneous space in finite times. Therefore, the inhomogeneous space in which the BP would equilibrate must be finite. In what follows, we will be using the above mentioned generalized stochastic noise strength to investigate the equilibrium of a BP in the inertial (non-overdamped) regime in an inhomogeneous space characterized by space dependent damping.

\section{generalized Langevin equation}
The generalized Langevin dynamics for a particle of mass $m$ in a space where the damping constant is coordinate dependent and the stochastic noise strength is as proposed in the previous section would look like
\ber
&& \dot{x} = v\\
&& m\dot{v} = -m\Gamma(x)v + F(x) + \Gamma(x)\sqrt{\frac{2mk_BT}{\left <\Gamma(x)\right >}}\eta(t).
\eer
Let us put Eq.7 and 8 to an additive noise form with the change of variables $u=v/\Gamma(x)$. We get equations
{
\ber
&&\dot{x} = \Gamma u\\
&&\dot{u} = -\Gamma u + \frac{F}{m\Gamma} -u^2\Gamma^\prime + \sqrt{\frac{2k_BT}{m\left <\Gamma(x)\right >}}\eta(t)\\
&&\dot{\Gamma} = \Gamma\Gamma^\prime u.
\eer  
In the above equations, a prime indicates spatial derivative, and $\Gamma$ is treated as a dynamical variable. The dynamics can always be cast into an additive noise form is the consequence of our consideration of noise strength as $\Gamma(x)\sqrt{2mk_BT/<\Gamma(x)>}$.
\par
The Fokker-Planck equation for the additive noise problem would now result from the continuity equation for the probability density which is of the form
\ber\nonumber
\frac{\partial P(x,u,\Gamma)}{\partial t} &=& -\frac{\partial }{\partial x}\left < \delta(x-x(t))\delta(u-u(t))\delta(\Gamma-\Gamma(x(t))\Gamma u \right > \\\nonumber
&-&\frac{\partial}{\partial u}\left < \delta(x-x(t))\delta(u-u(t))\delta(\Gamma-\Gamma(x(t))\left [-\Gamma u - \Gamma^\prime u^2 + \frac{F}{m\Gamma}+\sqrt{\frac{2k_BT}{m<\Gamma>}}\right ]\right >\\
&-& \frac{\partial}{\partial \Gamma}\left < \delta(x-x(t))\delta(u-u(t))\delta(\Gamma-\Gamma(x(t))\Gamma\Gamma^\prime u \right >,
\eer
where the probability density is $P(x,u,\Gamma)=<\delta(x-x(t))\delta(u-u(t))\delta(\Gamma-\Gamma(x(t))>$, and, the angular bracket indicates a noise average. The Fokker-Planck equation can now be readily written down as
\bea
u\frac{\partial}{\partial x}(\Gamma P) + \frac{F}{m\Gamma}\frac{\partial P}{\partial u} -\Gamma^\prime\frac{\partial}{\partial u}(u^2P) + u\Gamma^\prime\frac{\partial}{\partial \Gamma}(\Gamma P)
= \Gamma\left [ uP + \frac{k_BT}{m<\Gamma>\Gamma}\frac{\partial P}{\partial u}\right ],
\eea
where $P\equiv P(x,u,\Gamma)$ and, note that, $x$ and $u$ in the above equation are no longer the stochastic variables $x(t)$ and $u(t)$ where $\Gamma \equiv \Gamma(x)$ is an explicit function of $x$.
Now, taking the ansatz $P(x,u,\Gamma)\equiv P(\Gamma(x))P(x)P(u,\Gamma)=P(\Gamma(x))P(x)\exp{(-\frac{mu^2\Gamma<\Gamma>}{2k_BT})}$, the r.h.s. of the above equation becomes zero and the resulting equation is
\bea
u\frac{\partial \Gamma P(x,u,\Gamma)}{\partial x} + u\Gamma^\prime\frac{\partial \Gamma P(x,u,\Gamma)}{\partial \Gamma} +\frac{F}{m\Gamma}\frac{\partial P(x,u,\Gamma)}{\partial u} - \Gamma^\prime\frac{\partial u^2P(x,u,\Gamma)}{\partial u} = 0.
\eea
In what follows, we have to first differentiate with respect to $u$ and then replace $u$ by $v/\Gamma(x)$ to integrate out the $v$ dependence. Following this line of actions, after differentiating by $u$, we get
\ber\nonumber
\frac{\partial \Gamma P(x,u,\Gamma)}{\partial x} &+& \Gamma^\prime\frac{\partial \Gamma P(x,u,\Gamma)}{\partial \Gamma} -\frac{F<\Gamma>}{k_BT}P(x,u,\Gamma)-2\Gamma^\prime P(x,u,\Gamma)\\ &+& \frac{m\Gamma^\prime\Gamma<\Gamma>}{k_BT}u^2P(x,u,\Gamma)=0.
\eer
Now, first integrating out $v$ and then doing the other differentiations, we get
\ber\nonumber
&& \frac{3}{2}\Gamma^{1/2}\Gamma^\prime P(x)P(\Gamma(x)) + \Gamma^{3/2}P(\Gamma(x))\frac{\partial P(x)}{\partial x} + \Gamma^{3/2}P(x)\Gamma^\prime\frac{\partial P(\Gamma(x))}{\partial \Gamma} + \frac{3}{2}\Gamma^{1/2}\Gamma^\prime P(x)P(\Gamma(x))\\\nonumber &+& \Gamma^{3/2}P(x)\Gamma^\prime\frac{\partial P(\Gamma(x))}{\partial \Gamma} - \frac{F\Gamma^{1/2}<\Gamma>P(x)P(\Gamma(x))}{k_BT} - 2\Gamma^{1/2}\Gamma^\prime P(x)P(\Gamma(x))\\ &+& \Gamma^{1/2}\Gamma^\prime P(x)P(\Gamma(x)) =0,
\eer
where a common factor of $\sqrt{2\pi k_BT/m<\Gamma>}$ has been removed from each term. Following the standard procedure of finding the position distribution at this stage, of all the terms in the above equation, the second and the sixth terms when set together equal to zero gives 
\bea
P(x) = \exp{\left ( \frac{<\Gamma>}{k_BT}\int_\infty^x{dx^\prime\frac{F(x^\prime)}{\Gamma(x^\prime)}}\right )}.
\eea
Eq.16 with the rest of the terms reading as
\bea
\frac{dP(\Gamma)}{P(\Gamma)}=-\frac{d\Gamma}{\Gamma},
\eea
giving us
\bea
P(\Gamma(x)) = \frac{C}{\Gamma(x)},
\eea
where $C$ is a constant of integration.
\par
The full probability distribution at this stage looks like
\bea
P(x,v,\Gamma(x))\propto \frac{C}{\Gamma(x)} \exp{\left ( \frac{<\Gamma>}{k_BT}\int_\infty^x{dx^\prime\frac{F(x^\prime)}{\Gamma(x^\prime)}}\right )}\exp{\left (-\frac{mv^2<\Gamma>}{2\Gamma k_BT}\right )}.
\eea
The position distribution has been found out after integrating over the velocity $v$ to make it independent of the velocity distribution. Now, keeping in mind that the system is locally in some velocity state with a total probability unity, one must locally normalize the velocity distribution to finally get the probability distribution as
\bea
P(x,v,\Gamma(x)) = N\sqrt{\frac{m<\Gamma(x)>}{2\pi k_BT}} \frac{1}{\Gamma^{3/2}(x)} \exp{\left ( \frac{<\Gamma>}{k_BT}\int_\infty^x{dx^\prime\frac{F(x^\prime)}{\Gamma(x^\prime)}}\right )}\exp{\left (-\frac{mv^2<\Gamma>}{2\Gamma k_BT}\right )},
\eea
where $N$ is a normalization constant that absorbs $C$. The distribution of a free particle would have a modulated amplitude and would look like
\bea
P(u,\Gamma(x)) = N\sqrt{\frac{m<\Gamma(x)>}{2\pi k_BT}} \frac{1}{\Gamma^{3/2}(x)}\exp{\left (-\frac{mv^2<\Gamma>}{2\Gamma(x)k_BT}\right )}.
\eea
This is something special of the inhomogeneous space that the position distribution in equilibrium is not constant over space in the absence of a force. This is because of the fact that the removal of force is not removing inhomogeneity of space in the presence of other sources. The distribution must be inhomogeneous to have a local velocity distribution with zero average as is evident from the present analysis.
}

\begin{figure}
\subfigure []
{\includegraphics[width= 8 cm,angle=0]{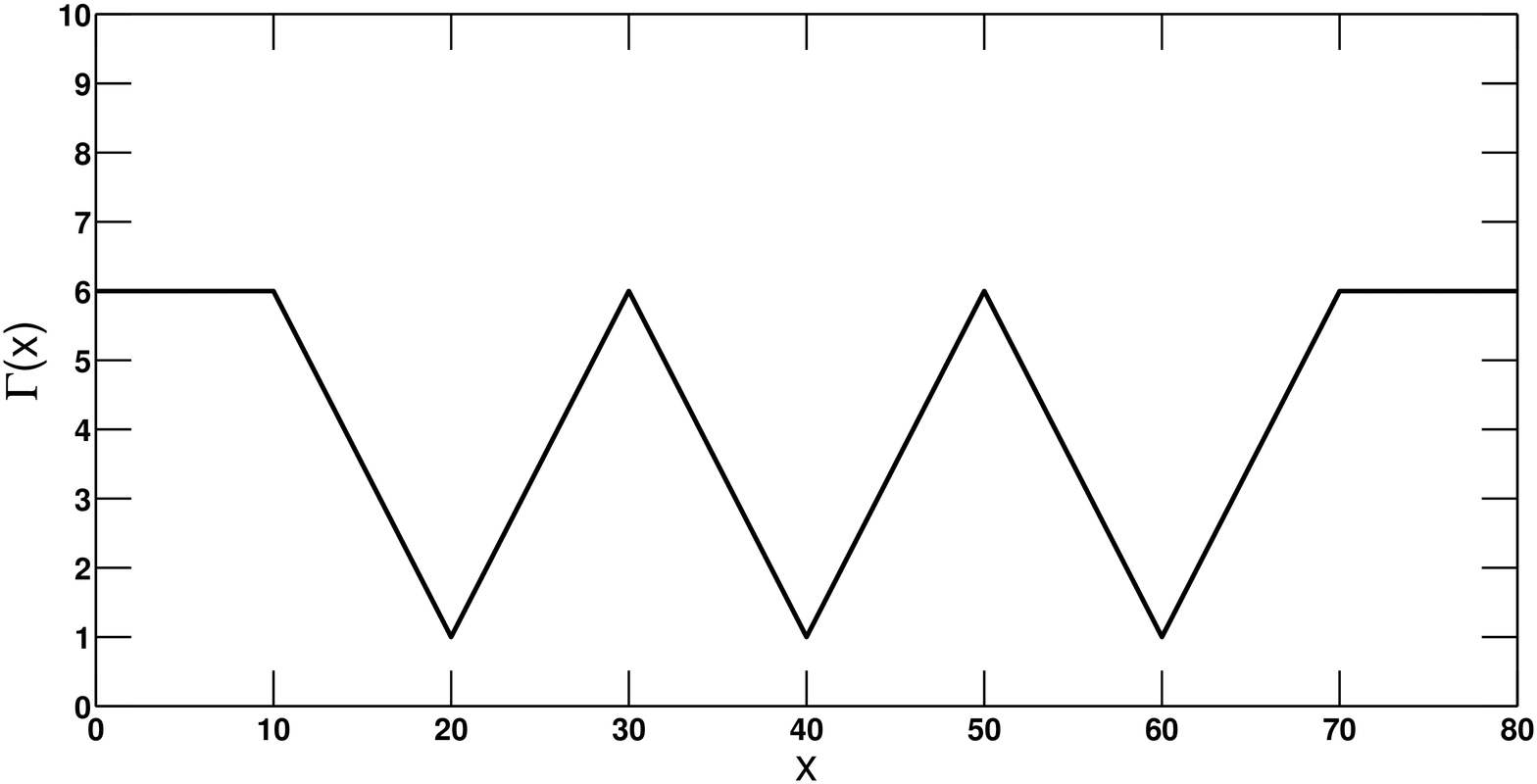}
\label{Fig:a}}
\subfigure []
{\includegraphics[width= 8 cm,angle=0]{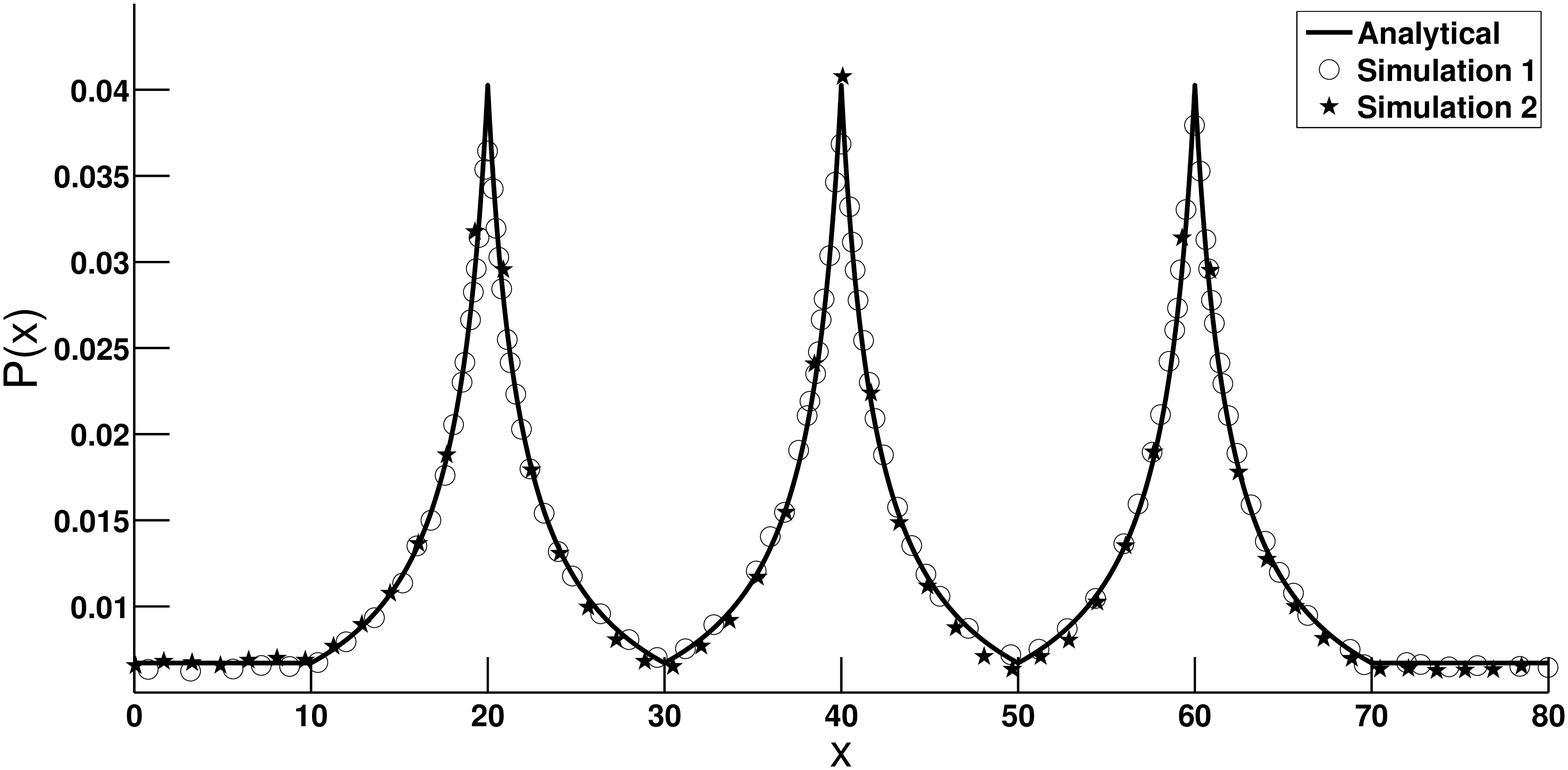}
\label{Fig:b}}
\subfigure []
{\includegraphics[width= 8 cm,angle=0]{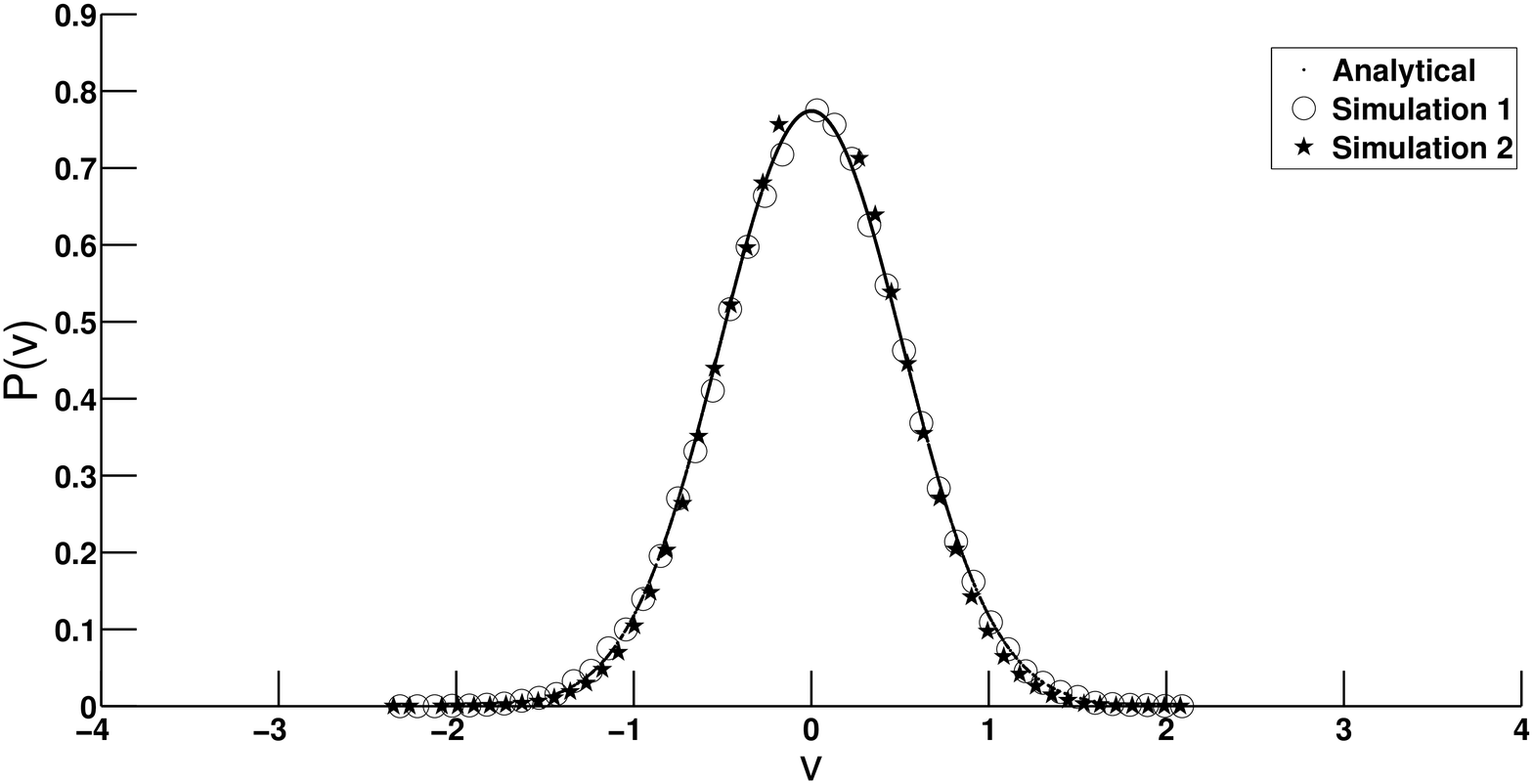}
\label{Fig:c}}
\subfigure []
{\includegraphics[width= 8 cm,angle=0]{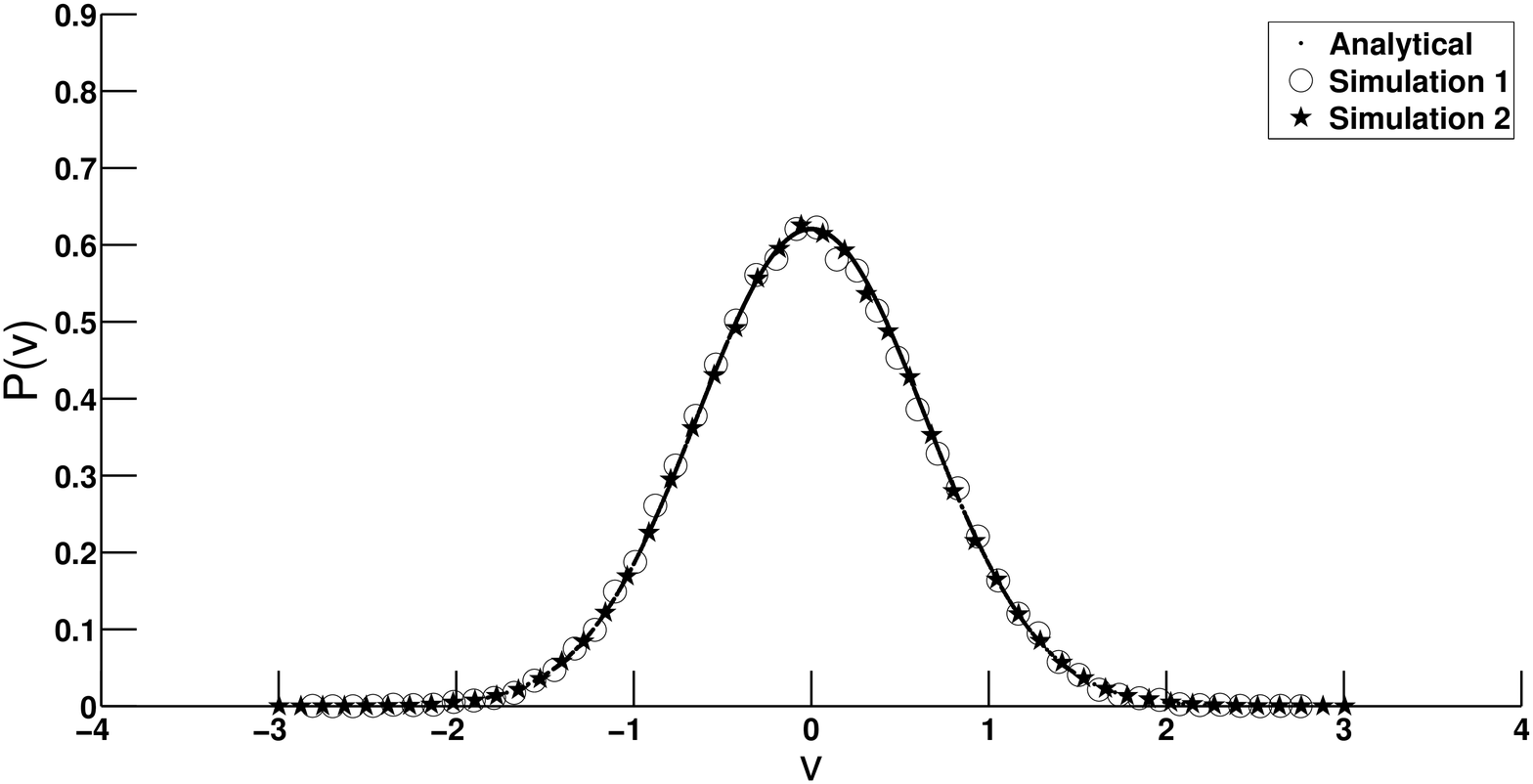}
\label{Fig:d}}
\caption []{{Free particle with periodic boundary condition:} (a) Plot of $\Gamma(x)$ vs $x$, (b) Plot of $P(x)$ vs $x$, (c) Plot of $P(v)$ vs $v$ at $x=42$ within a bin size $\bigtriangleup x = 0.01$ and (d)Plot of $P(v)$ vs $v$ at $x=47.5$ within a bin size $\bigtriangleup x = 0.01$. }
\end{figure}

%\begin{figure}
%\subfigure []
%{\includegraphics[width= 8 cm,angle=0]{Fig_2a.eps}
%\label{Fig:a}}
%\subfigure []
%{\includegraphics[width= 8 cm,angle=0]{Fig_2b.eps}
%\label{Fig:b}}
%\subfigure []
%{\includegraphics[width= 8 cm,angle=0]{Fig_2c.eps}
%\label{Fig:c}}
%\subfigure []
%{\includegraphics[width= 8 cm,angle=0]{Fig_2d.eps}
%\label{Fig:d}}
%\caption []{{Free particle with hard wall boundary condition:} (a) Plot of $\Gamma(x)$ vs $x$, (b) Plot of $P(x)$ vs $x$, (c) Plot of $P(v)$ vs $v$ at $x=38$ within a bin size $\bigtriangleup x = 0.01$ and (d)Plot of $P(v)$ vs $v$ at $x=46.25$ within a bin size $\bigtriangleup x = 0.01$. }
%\end{figure}

%\begin{figure}
%\subfigure []
%{\includegraphics[width= 8 cm,angle=0]{Fig_3a.eps}
%\label{Fig:a}}
%\subfigure []
%{\includegraphics[width= 8 cm,angle=0]{Fig_3b.eps}
%\label{Fig:b}}
%\subfigure []
%{\includegraphics[width= 8 cm,angle=0]{Fig_3c.eps}
%\label{Fig:c}}
%\subfigure []
%{\includegraphics[width= 8 cm,angle=0]{Fig_3d.eps}
%\label{Fig:d}}
%\caption []{{Particle in a harmonic trap:} (a) Plot of $\Gamma(x)$ vs $x$, (b) Plot of $P(x)$ vs $x$, (c) Plot of $P(v)$ vs $v$ at $x=82.5$ within a bin size $\bigtriangleup x = 0.01$ and (d)Plot of $P(v)$ vs $v$ at $x=95$ within a bin size $\bigtriangleup x = 0.01$.}
%\end{figure}

\begin{figure}
\subfigure []
{\includegraphics[width= 8 cm,angle=0]{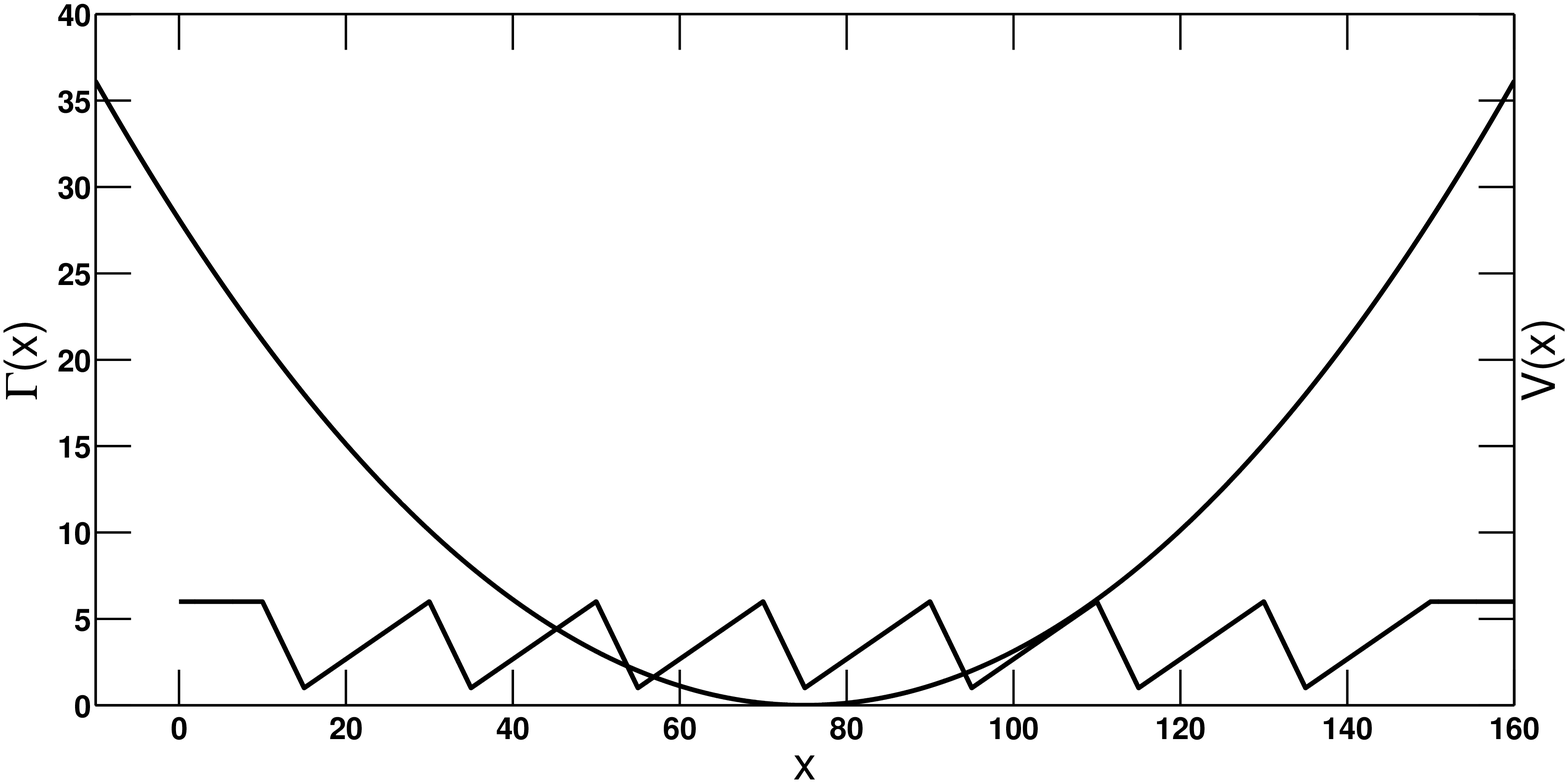}
\label{Fig:a}}
\subfigure []
{\includegraphics[width= 8 cm,angle=0]{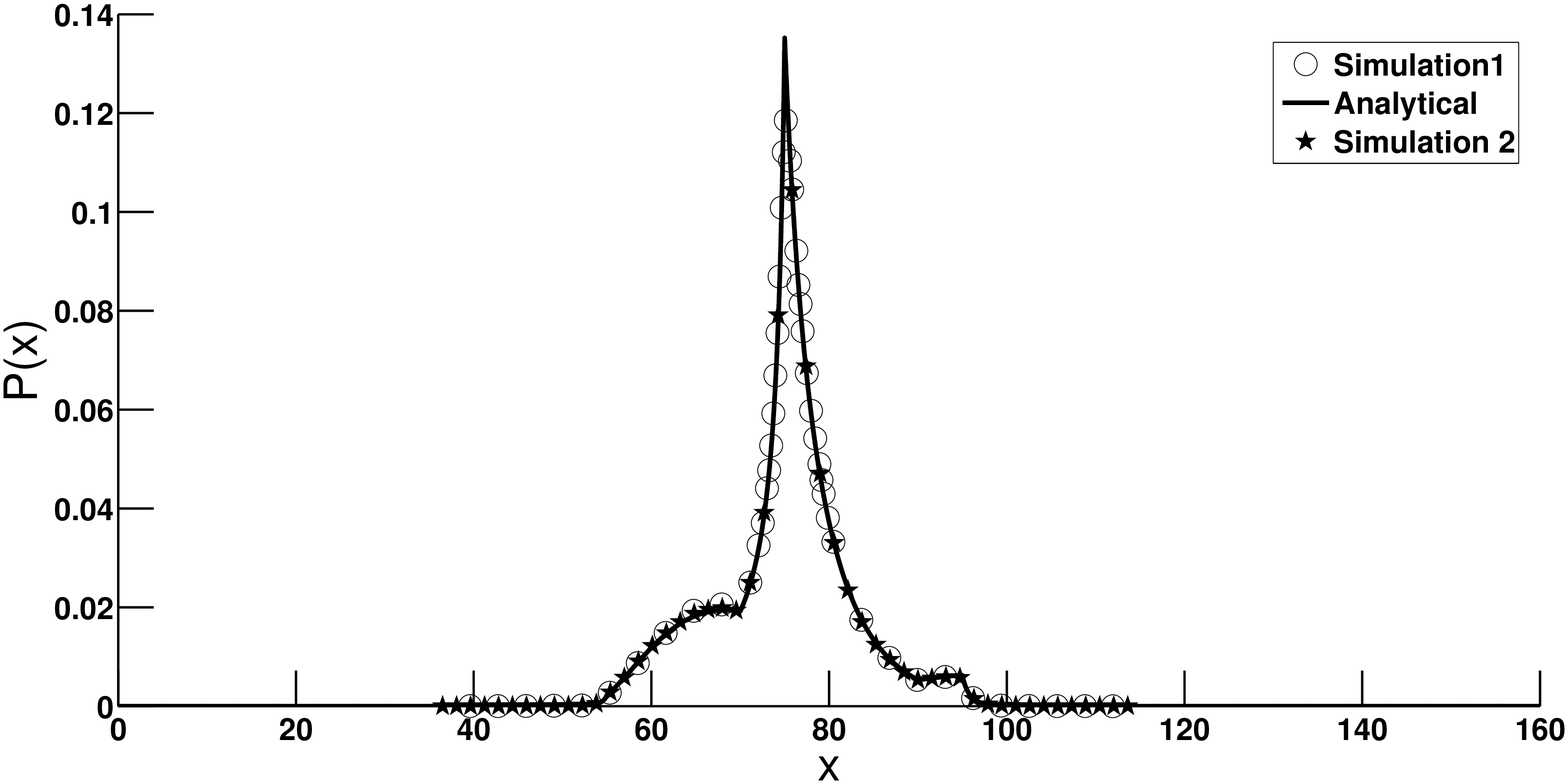}
\label{Fig:b}}
\subfigure []
{\includegraphics[width= 8 cm,angle=0]{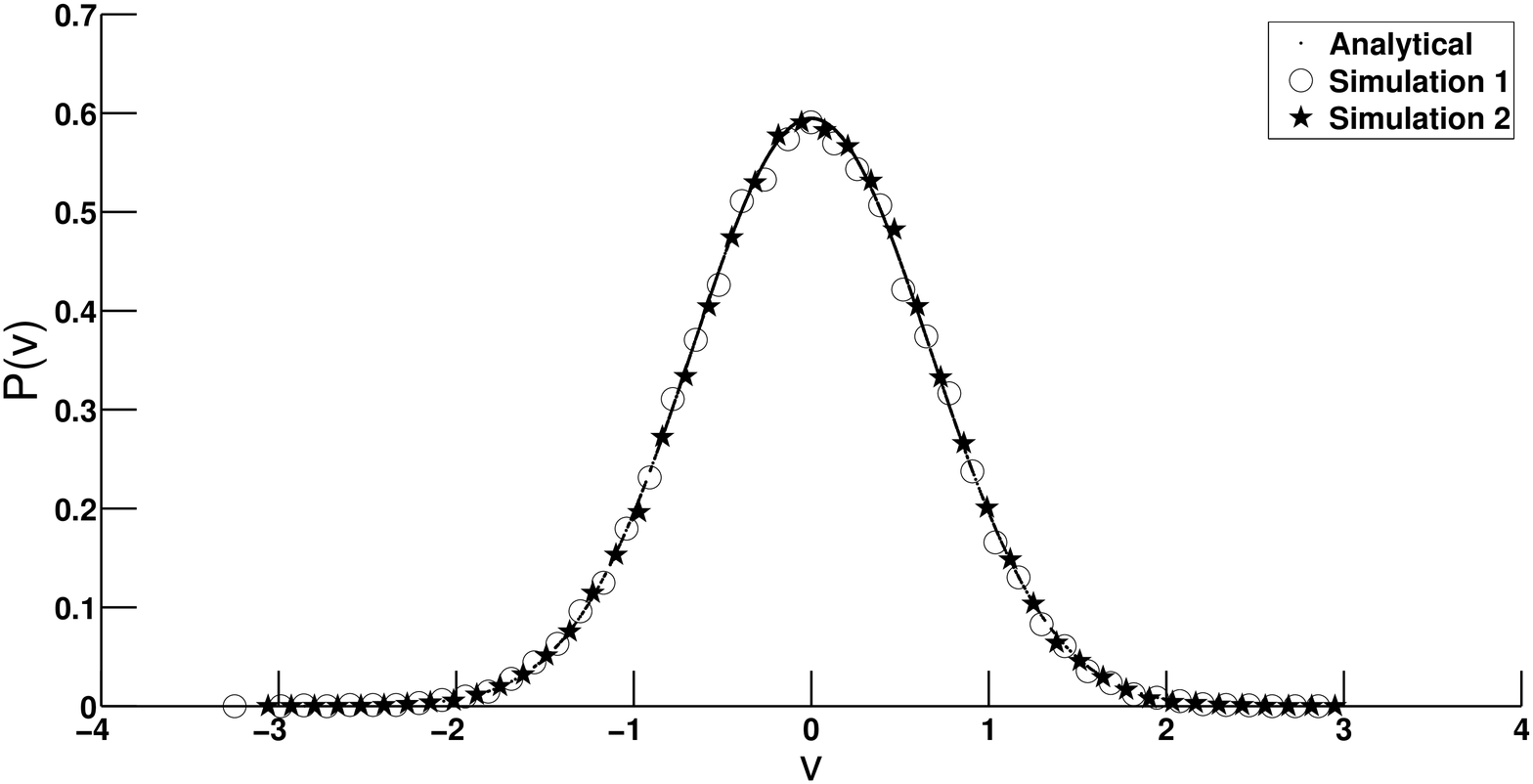}
\label{Fig:c}}
\subfigure []
{\includegraphics[width= 8 cm,angle=0]{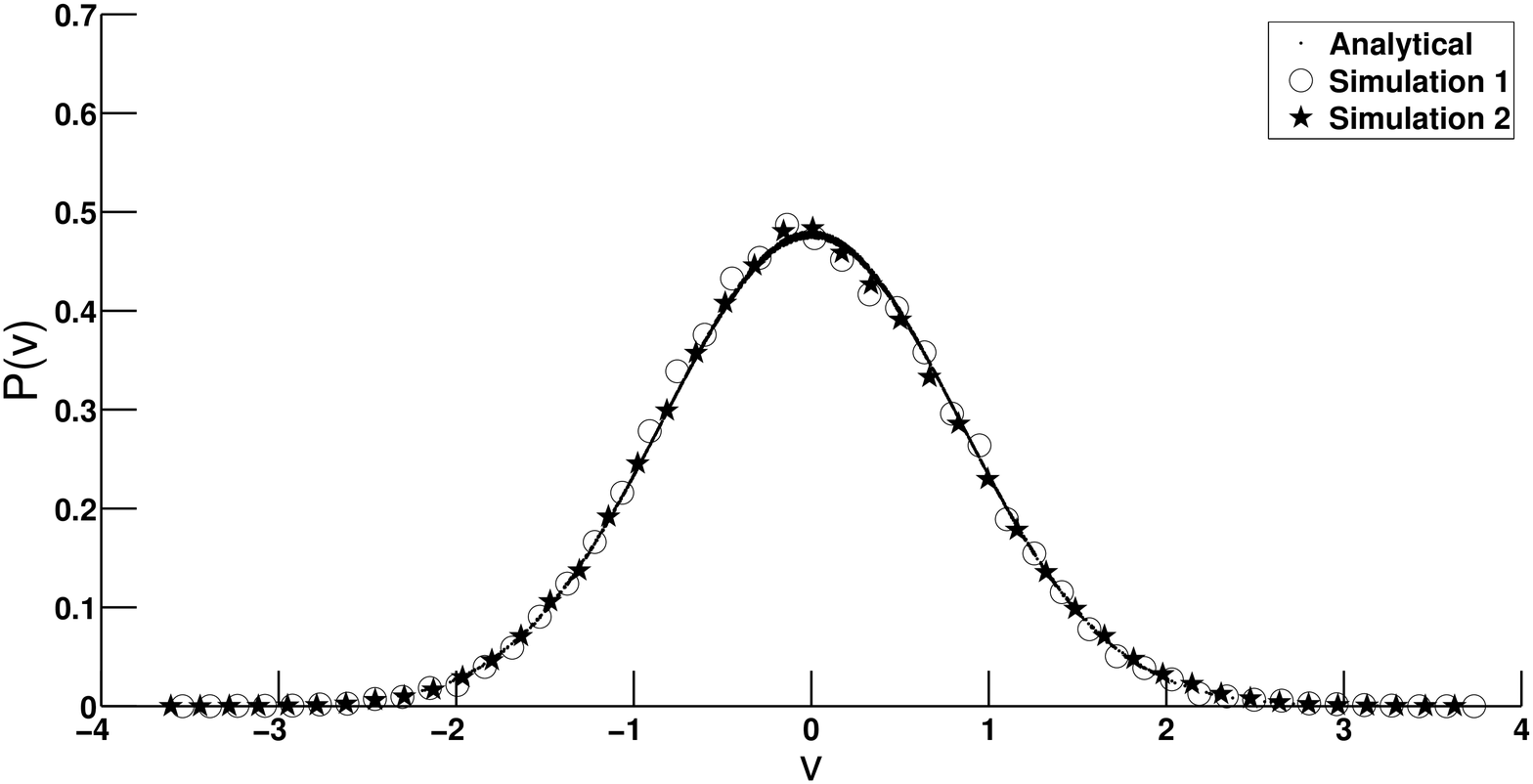}
\label{Fig:d}}
\caption []{{Particle in harmonic trap:} (a) Plot of $\Gamma(x)$ vs $x$, (b) Plot of $P(x)$ vs $x$, (c) Plot of $P(v)$ vs $v$ at $x=78.75$ within a bin size $\bigtriangleup x = 0.01$ and (d)Plot of $P(v)$ vs $v$ at $x=92.5$ within a bin size $\bigtriangleup x = 0.1$.}
\end{figure} 

\section{Numerical results}
For the numerical investigation of the dynamics of a BP, we consider two profiles of $\Gamma(x)$, one is inversion-symmetric (profile 1) with a periodic boundary condition and the other is an inversion-asymmetric form (profile 2) within a global harmonic potential. We have done two sets of numerical simulations. The first one is numerical integration of Eq.7 and 8 using the noise strength corresponding to the position of the particle just at the beginning of each discrete time interval (It\^o) which we mention as {\it simulation 1}. In the second case, we have numerically integrated Eq.9-11 which we mention as {simulation 2} in the figures. We compare the result of these two simulations with theoretical ones. In Fig.1, we have plotted the position and velocity distributions for a free particle with a inversion-symmetric $\Gamma(x)$. In each of the figures, the sub-figure $a$ shows the actual $\Gamma(x)$ profile, where the sub-figure $b$ shows a plot of position distribution. Sub-figures $c$ and $d$ are showing the locally normalized velocity distributions at two different points over the accessible space. Fig.2 has been plotted for a particle confined in a harmonic trap (spring constant $0.01$) with an asymmetric $\Gamma(x)$ profile. The scheme of the sub-plots are the same as that of Fig.1. Mass of the particles has throughout been considered to be unity for the sake of simplicity. We have also done simulations of the free particle with asymmetric damping (profile 2) and harmonically confined particle with symmetric damping (profile 1) and results are as consistent with theory as are shown for the reverse cases here. 
\par
In the simulation, an ensemble of 1000 particles is considered and equations are evolved for $4\times10^4$ time steps of size $\bigtriangleup t=10^{-3}$. We waited for $2\times10^4$ steps for the system to equilibrate before extracting any data. In the simulation a value of $\delta = k_BT/<\Gamma(x)>$ is set by hand. For graphs in Fig.1, $\delta = 0.118$ and for the Fig.2 $\delta = 0.2$. We also varied $\delta$ over many values and saw similar results. Various values set to this quantity are given in tabular form for simulation 1 and 2. One can directly evaluate the $<\Gamma(x)>$ numerically from the profiles set and the numerically obtained probability distribution. Then, using this value of $<\Gamma(x)>$ one can extract the value of $k_BT$ from the $\delta$. A comparison of $k_BT$ with the average squared velocity of the unit mass particle is given in tabular form below. Note that, this is an average $v^2$ over the whole inhomogeneous space. The average has been done on $1000$ trajectories of length $5000$ discrete time steps.  
\par
While doing the numerical simulation of Eq.7 and 8 and evaluating the noise term, we kept the noise strength to its value at the beginning of each interval $dt$ for every discrete update. This is essentially an It\^o convention to adopt. The result is in good agreement with that obtained from Eq.9-11 which is a mapping to an additive noise equivalent dynamics. Both the simulations are in good agreement with the theoretical results as well. Note that, it has been possible to map the dynamics to an additive noise equivalent because of the use of noise strength in the form we have proposed in Ref.\cite{genfdt}. We claim the resulting distribution to be equilibrium probability distribution of the generalized (with mass) Langevin dynamics. The local Maxwellian distribution of velocity can be considered as the evidence in support of our claim. Being able to map a multiplicative noise problem to an additive noise form can be considered to be the prerequisite to have equilibrium in the system. If that is true, then, there exists no dilemma of conventions in dealing with the Langevin dynamics of stochastic systems which would equilibrate.

\begin{table}[ht]
\caption{simulation 1}
\centering
\begin{tabular}{c*{5}{c}}
Profile           & $\delta$ & $<\Gamma>$ & $k_BT$ & $<v^2>$ & particle type \\
\hline
1                 & 0.118 & 3.2214 & 0.3801 & 0.3788 & free   \\
2                 & 0.118 & 3.2214 & 0.3801 & 0.3833 & free  \\
1                 & 0.20  & 2.7885 & 0.5577 & 0.5592 & harmonic \\
2                 & 0.20  & 2.7998 & 0.5600 & 0.5647 & harmonic \\

\end{tabular}
\end{table}

\begin{table}[ht]
\caption{simulation 2}
\centering
\begin{tabular}{c*{5}{c}}
Profile           & $\delta$ & $<\Gamma>$ & $k_BT$ & $<v^2>$ & particle type \\
\hline
1                 & 0.118 & 3.2214 & 0.3801 & 0.3793 & free   \\
2                 & 0.118 & 3.2214 & 0.3801 & 0.3846 & free  \\
1                 & 0.20  & 2.7885 & 0.5577 & 0.5569 & harmonic \\
2                 & 0.20  & 2.7998 & 0.5600 & 0.5568 & harmonic \\

\end{tabular}
\end{table}

\section{Discussion}
The main result of our present analysis in support of our previous work \cite{genfdt} is to show that, the Maxwell-Boltzmann distributions have to be modified to be an equilibrium distribution over an inhomogeneous space where the inhomogeneity is due to a spatial dependence of the damping constant. This type of a spatial inhomogeneity is ubiquitous in complex molecules like polymers, proteins and colloids. For example, in the case of a polymer, the proximity of other monomers to a particular one can always alter the damping seen by the particular monomer just in the same way as the damping constant of a Brownian particle changes near a wall \cite{lan}. Thus, the internal (conformation) space of such complex molecules is a good and general representation of the kind of inhomogeneous space considered in the present context. Such situations are very common, and we are proposing an alternative distribution for equilibrium of such systems.
\par
The probability distribution that we have got in the present case for a free particle, in particular, is not uniform in space. However, in the over-damped case of a free particle \cite{genfdt}, the probability distribution is uniform over space in the absence of the force field $F(x)$. Such a result is expected when the velocity is practically absent in the over-damped dynamics. But, in the presence of the inertial term, velocity distribution being a local function of space, the probability of finding a particle at different positions cannot be uniform because that would lead to existence of currents. The variation of the width of the velocity distribution with the $\Gamma(x)$ is evident from the sub-figures c and d. Particles are more probable at positions where they have a lesser width of the velocity distribution than at those places where the distribution has bigger width. This picture is completely lost as soon as one enforces a Maxwellian velocity distribution uniformly throughout the space without having any coupling to the inhomogeneities created by sources other than a conservative force field. In the context of free particles, this is the striking feature of the present analysis as opposed to the standard ones. Note that, in the presence of a force, the exponential dependence of the probability distribution on the force being too strong, this spatial dependence of the amplitude gets somehow masked. So, for an experimental verification of our present results, the probability distribution of a free particle is more desirable than a bound one to a force field.  
\par
{
Let us try to qualitatively relate the free particle distribution of ours to give us a novel clue as to what might happen in the case of a folding protein which is an unsolved problem till date. The Levinthal's paradox, in the context of protein folding, is about understanding how a protein folds so quickly given so many equivalent configurations and very rugged energy landscape. If one considers all the conformational states to be equivalent in the absence of Boltzmann selection, then, there are too many states to sample to get to the desired one. Alternatively, when the Boltzmann distribution of states are considered, the energy landscape becomes so rugged that the dynamics becomes extremely slow and glassy. This is the general problem and one tries to find out how a path is cut over the rugged energy landscape so that the system quickly moves to its global energy minimum which many believe to represent the native fold. The Maxwell-Boltzmann distribution has so far not been satisfactory, so far, in predicting such a preferred path through the rugged energy landscape. 
\par
On the contrary, our probability distribution strongly depends on $\Gamma(x)$, even for a free particle with mass. One can speculate here that $\Gamma(x)$ can cut a preferred path over the rugged energy landscape when the probability distribution is of the form as is shown here. But, how exactly it could do that requires a knowledge of this phenomenological constant. Let us take note of the fact that protein folding transitions are generally accompanied by an early collapse transition driven by hydrophobic forces. It's not unreasonable to think that, following the collapse, the system reaches a state where the damping is minimal due to the reduction of the accessible surface area to the bath degrees of freedom. The dynamics no longer remains over-damped and the inertial terms should become important. In such a regime, the dynamics not only becomes fast, the probability distribution in such a situation would be the $P(\Gamma(x),v)$ of a free particle which is strongly dependent on the inverse of the small $\Gamma(x)$. This strong dependence of the probability distribution on $\Gamma(x)$ indicates that the dynamics is predominantly confined to the conformations with minimal $\Gamma(x)$ which possibly cuts out the preferred path to the global minimum in the presence of the other structural constraints on bending and translations. We would like to systematically look at these scenarios in future on the basis of our formalism.
}

\begin{acknowledgements} 

\end{acknowledgements}

\end{document}